\documentclass[twocolumn,preprintnumbers,amsmath,amssymb,aps,prb]{revtex4}
\usepackage{graphicx}
\begin{document}

\title{Avalanches and Criticality in Driven Magnetic Skyrmions}
 
\author{S. A. D{\' \i}az$^{1,2}$, C. Reichhardt$^{1}$, D. P. Arovas$^{2}$, A. Saxena$^{1}$ and C. J. O. Reichhardt$^{1}$}
\affiliation{$^{1}$Theoretical Division and Center for Nonlinear Studies,
Los Alamos National Laboratory, Los Alamos, New Mexico 87545, USA}
\affiliation{$^{2}$Department of Physics, University of California San Diego, La Jolla, 92093, USA}

\date{\today}

\begin{abstract}
We show using numerical simulations that slowly driven skyrmions interacting with random 
pinning move
via correlated jumps or avalanches.  
The avalanches exhibit power law distributions in their duration and size, 
and the average avalanche shape for different avalanche durations can    
be scaled to a universal function, in agreement with theoretical predictions 
for systems in a  nonequilibrium critical state. A distinctive feature
of skyrmions is the influence of the
non-dissipative Magnus term.
When we increase
the ratio of the Magnus term to the damping term,
a change in the universality class of the behavior occurs, the
average avalanche shape becomes
increasingly asymmetric,
and individual avalanches exhibit motion in the direction
perpendicular to their own density gradient.    
\end{abstract}

\maketitle

\vskip 2pc

{\it Introduction--}
Magnetic skyrmions are nanoscale particlelike
spin textures that were first observed in chiral magnets in 2009 \cite{1,2} 
and have since been identified in a growing variety of materials,
including several that 
support skyrmions at room temperature \cite{3,4,5,6,7,8}.
Skyrmions can exhibit depinning phenomena under an applied
current \cite{2,9,10,11,12,13,14,15,16},
and their ability to be set in motion along with their size scale 
make them promising candidates for a variety of applications \cite{17,18}.
A key feature of skyrmions
that is distinct from other depinning systems \cite{19} is
the strong influence on the skyrmion motion of the non-dissipative Magnus
term, which arises from the skyrmion topology \cite{2}.
Strong Magnus terms are also relevant for vortex depinning in
neutron star crusts \cite{34,35}.
In particle-based models of
vortices in type-II superconductors or colloidal particles,
the motion is dominated by the damping term which aligns the
particle velocity with the external forces \cite{19}.
In contrast, 
the Magnus term aligns the particle velocity perpendicular to the direction of the
external forces, causing the skyrmions to move at an angle called the
intrinsic skyrmion Hall angle $\theta^{\rm int}_{Sk}$ with respect to the external forces
\cite{2,10,12,13,14}.
As recently shown,
the Magnus term strongly affects
the overall skyrmion dynamics in the presence of disorder,
with 
the measured skyrmion Hall angle
starting at zero or a small value for drives just above depinning and
gradually
increasing to the intrinsic or pin-free $\theta^{\rm int}_{Sk}$ value as the
drive increases and the skyrmions
move faster \cite{12,13,14,15,16,20,21,22,23}. 

In many slowly driven systems with quenched disorder, 
the motion near depinning takes the form of bursts or avalanches
of the type
observed in
driven magnetic domain walls \cite{24,24a,25}, 
vortices in type-II superconductors \cite{19,26}, earthquake models \cite{28}, 
and near yielding transitions in sheared materials \cite{29,30}.  
Avalanches or
so-called crackling 
noise arise in a wide range of
collectively interacting driven systems, 
and scaling properties of
the avalanche size distributions as well as
the average avalanche shape can be used 
to determine whether the system is at 
a nonequilibrium critical point and to identify its universality class
\cite{31,32,33}. 
In many avalanche systems, the dynamics is
overdamped, but
when non-dissipative effects
become important,
the statistics of the  avalanches can 
change.
In particular,
the average avalanche shape
becomes asymmetric in the presence of an effective mass or
stress overshoots
\cite{25,32,33}.
An open question is whether skyrmions
can exhibit avalanche dynamics and,
if so, what impact the Magnus term would have on such dynamics.
It 
is important to understand intermittent skyrmion dynamics 
near the depinning threshold in order
to fully realize applications which require skyrmions to be moved and
stopped in a controlled fashion, such as in skyrmion race track memories \cite{18}.    

In this work we numerically examine avalanches of slowly driven skyrmions
moving over quenched disorder
for varied ratios $\alpha_m/\alpha_d$ of the
Magnus term to the damping term. 
When $\alpha_m/\alpha_d \leq 1.73$,
corresponding to intrinsic skyrmion Hall angles of $\theta^{\rm int}_{Sk} \leq 60^0$,
the skyrmion avalanches are
power law distributed in both size and duration,
and the average avalanche shape for a fixed duration
can be scaled to a   
universal curve as predicted for systems  
in a nonequilibrium critical state \cite{31,32,33}.
For larger values of the Magnus term,
the avalanches develop a characteristic size and
the average avalanche shape becomes strongly asymmetric,
indicative of an effective negative mass similar to that observed
for avalanche distributions in certain domain wall systems \cite{25}. 

\begin{figure}
\includegraphics[width=3.5in]{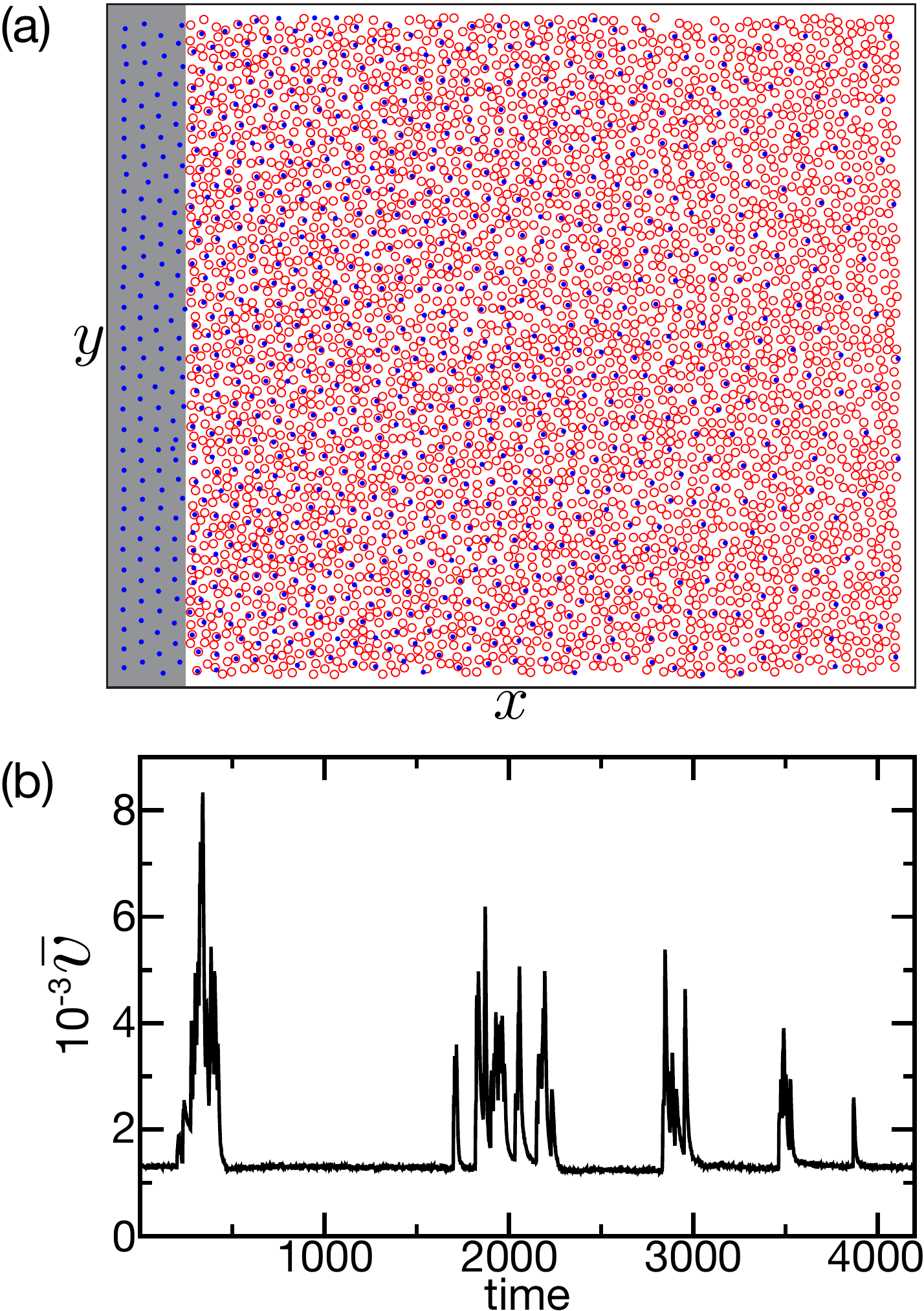}
\caption{(a) Snapshot of the system showing the skyrmions (solid dots) 
  and pinning sites (open circles).  Skyrmions are introduced in the unpinned region
  on the left side of the sample and removed when they reach the right side of the
  sample.
  Once the system reaches a steady state,
  individual skyrmions are added at a slow rate.
  (b) A segment of the time series of the net skyrmion velocity,
  $\bar{v}$, versus time in simulation time steps.
  Clear skyrmion avalanche events appear.
}
\label{fig:1}
\end{figure}

{\it Simulation and System---}
In Fig.~\ref{fig:1} we show a snapshot of our 2D system
which has periodic boundary conditions only in the $y$-direction and contains
$N$ skyrmions interacting with $N_p$ randomly placed pinning sites.
The skyrmions are modeled as particles with
dynamics  governed by the modified Thiele equation,
used previously to study skyrmions interacting with
random
\cite{12,16} and periodic
\cite{21,22} pinning substrates. 
The equation of motion of a single skyrmion $i$ is: 
\begin{equation}  
\alpha_d {\bf v}_{i} - \alpha_m {{\bf \hat z}} \times {\bf v}_{i} = 
      {\bf F}^{ss}_{i} + {\bf F}^{sp}_{i} .
\end{equation}
Here 
${\bf r}_i$ is the skyrmion position 
and 
${\bf v}_{i} = {d {\bf r}_{i}}/{dt}$ is the skyrmion velocity. 
The damping constant is  $\alpha_d$ while
$\alpha_m$ is the strength of the Magnus term. In the absence of pinning,
a skyrmion experiencing a uniform external force
moves  at the intrinsic skyrmion
Hall angle of $\theta^{\rm int}_{Sk} = \tan^{-1}(\alpha_{m}/\alpha_{d})$
with respect to the direction of the external force,
and in the overdamped limit of $\alpha_{m}  = 0$, $\theta^{\rm int}_{Sk} = 0^{\circ}$.  
The skyrmion-skyrmion repulsive interaction force is given by
${\bf F}^{ss}_{i} = \sum^{N}_{j=1} K_{1}(r_{ij}) \hat{\bf r}_{ij}$
where $r_{ij}=|{\bf r}_i - {\bf r}_j|$,
$\hat{\bf r}_{ij}=({\bf r}_i - {\bf r}_j)/r_{ij}$, and
$K_{1}$ is a modified Bessel function.
The pinning force from the quenched disorder ${\bf F}^{sp}_{i}$
arises from $N_p$ randomly placed non-overlapping
harmonic traps with maximum pinning force 
$F_{p}$ and radius $R_{p} = 0.15$.
The system dimensions are
$L_{x} = 26$ and $L_{y} = 24$, and there is a pin-free region
extending from $x=0$ to $x=4$.  An artificial wall of stationary skyrmions is
placed to the left of $x=0$ to provide confinement.
The skyrmions are driven by a gradient, introduced by slowly dropping
skyrmions into the pin-free region and allowing them to move into the pinned
region under the force of their mutual repulsion\cite{myaval}.   Skyrmions that reach the
right edge of the sample are removed from the simulation.
After the system reaches a steady state,
which typically requires
$2 \times 10^3$ skyrmion drops,
we examine individual avalanches by
measuring the net skyrmion velocity response
$\bar v=N^{-1}\sum_{i=0}^{N}|{\bf v}_i|$
between drops, as illustrated in
Fig.~\ref{fig:1}(b).  We drop skyrmions at a slow enough rate
that the time series $\bar v(t)$ shows 
well-defined avalanches separated by intervals of no motion.
We consider five different intrinsic skyrmion Hall angles
$\theta^{\rm int}_{Sk} = 0^{\circ}$, $30^{\circ}$, $45^{\circ}$, $60^{\circ}$,
and $80^{\circ}$, where we
fix $\alpha_{d} = 1.0$ and vary $\alpha_{m}$.
We
studied
several different pinning densities and strengths,
but here we focus on
systems with $N_{p} = 3700$ and $F_{p} = 1.0$.
Experimentally our system
corresponds to skyrmions
entering from the edge of a sample or
moving from a pin-free to a pinned region of the sample driven by a
slowly changing magnetic field or small applied current.

\begin{figure}
\includegraphics[width=3.5in]{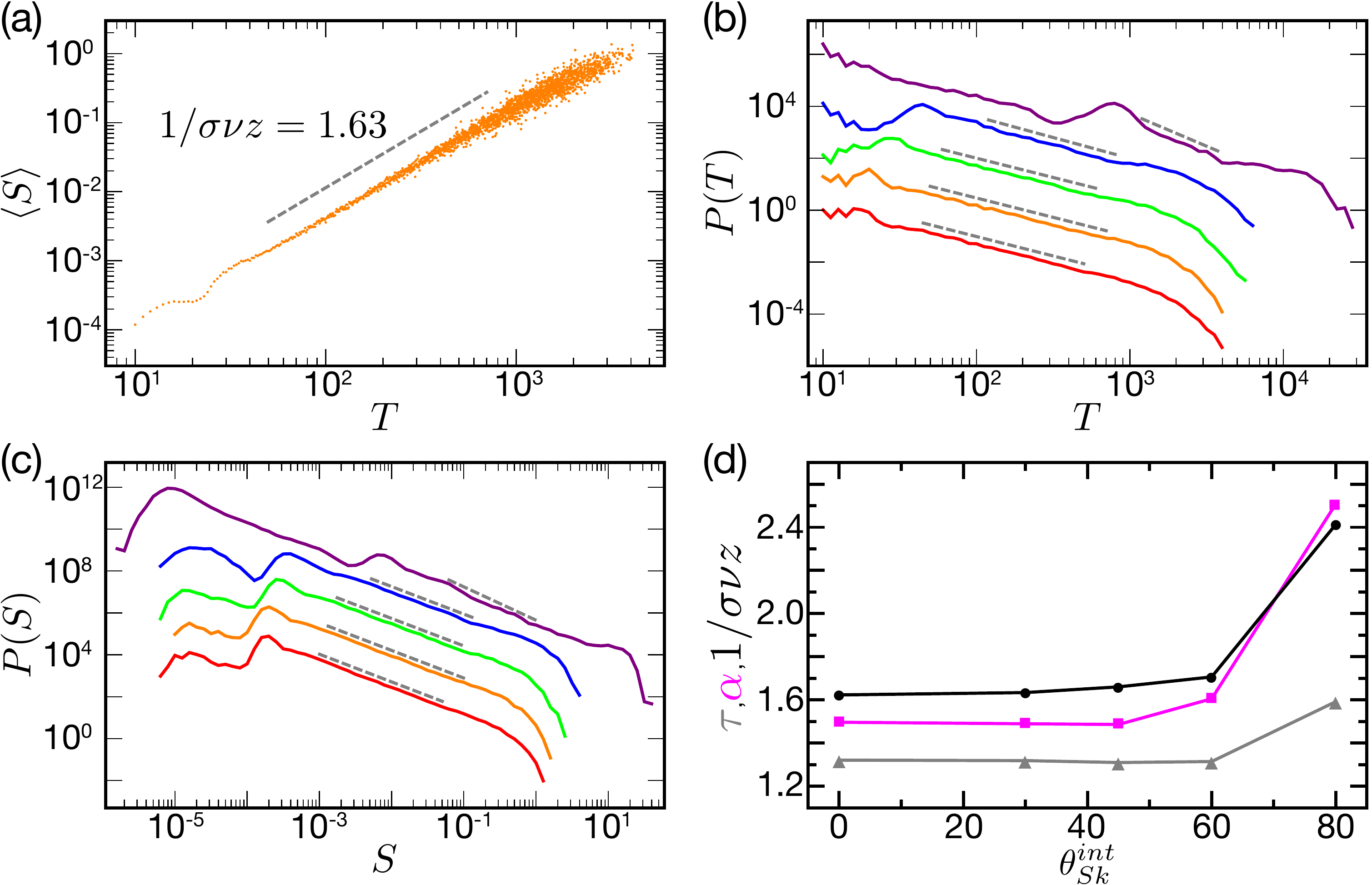}
\caption{(a) Average avalanche size
  $\langle S\rangle$ vs avalanche duration $T$
  for $\theta^{\rm int}_{Sk} = 30^\circ$. 
Dashed line is a fit to
$\langle S\rangle \propto T^{1/\sigma\nu z}$  with $1/\sigma\nu z = 1.63$.
(b) Distribution of avalanche durations $P(T)$ and
(c) distribution of avalanche sizes $P(S)$
for $\theta^{\rm int}_{Sk} = 0^\circ$, $30^\circ$, $45^\circ$,
$60^\circ$, and $80^\circ$, from bottom to top.  The curves have been shifted
vertically for clarity.
(d) The scaling exponents $\tau$ (triangles), $\alpha$ (squares),
and $1/\sigma\nu z$ (circles)
vs $\theta^{\rm int}_{Sk}$.   For $\theta^{\rm int}_{Sk} \leq 60^\circ$, 
$\alpha = 1.5$, $\tau = 1.33$, and $1/\sigma\nu z = 1.63$, while
for $\theta^{\rm int}_{Sk} = 80^\circ$, $\alpha = 2.5$,
$\tau = 1.6$, and $1/\sigma\nu z = 2.4$.    
}
\label{fig:2}
\end{figure}

{\it Results---}
From the time series of the skyrmion velocity ${\bar v}(t)$ we determine
the avalanche duration $T$ as the time during which ${\bar v}>v_{th}$, where
$v_{th}$ is a threshold velocity.
We define the avalanche size $S$ as the time integral
$S = \int_{t_0}^{t_0+T} {\bar v}(t) - v_{th}$ over the duration of the avalanche.
Near a critical point, various quantities associated with
the avalanches are expected to scale as power laws \cite{31}:
the average avalanche size $\langle S\rangle(T) \propto T^{1/\sigma\nu z}$,
the distribution of avalanche durations 
$P(T) \propto T^{-\alpha}$,
and the avalanche size  distribution $P(S)\propto S^{-\tau}$.
In Fig.~\ref{fig:2}(a) we plot $\langle S\rangle$ versus $T$
for a system with $\theta^{\rm int}_{Sk} = 30^\circ$ , while
Figs.~\ref{fig:2}(b,c) show the corresponding
$P(T)$ and $P(S)$ for $\theta^{\rm int}_{Sk} = 0^\circ$ 
to $80^{\circ}$.  In each case we find a range of power law scaling.
In Fig.~\ref{fig:2}(d) we plot the extracted critical exponents 
$\tau$, $\alpha$, and $1/\sigma\nu z$ versus $\theta^{\rm int}_{Sk}$.
The exponents are roughly constant for 
$\theta^{\rm int}_{Sk} \leq 60^{\circ}$
with $\tau = 1.33$, $\alpha = 1.5$, and $1/\sigma\nu z = 1.63$.
For $\theta^{\rm int}_{Sk} = 80^{\circ}$, we find longer avalanches of larger size,
as indicated by the changes in $P(T)$ and $P(S)$,
while $P(T)$ develops a maximum due to the emergence of a characteristic
avalanche size.
If we consider only
the larger avalanches from the $\theta^{\rm int}_{Sk} = 80^\circ$ sample, we obtain
considerably larger exponents
of $1/\sigma\nu z  = 2.4$, $\alpha \approx 2.5$, and   $\tau = 1.6$,    
as shown in Fig.~\ref{fig:2}(d).
The exponents for a system in a critical state are predicted to 
satisfy the following relation\cite{31}:
\begin{equation}
\frac{\alpha -1}{\tau -1} = \frac{1}{\sigma\nu z} .
\end{equation} 
Samples with $\theta^{\rm int}_{Sk} < 60^{\circ}$ obey this relation,
samples with
$\theta^{\rm int}_{Sk} = 60^{\circ}$ give $1.55$ for the right hand side and 1.63 for the
left hand side, and samples
with $\theta = 80^{\circ}$ again obey this relation.

\begin{figure}[h!]
\includegraphics[width=3.5in]{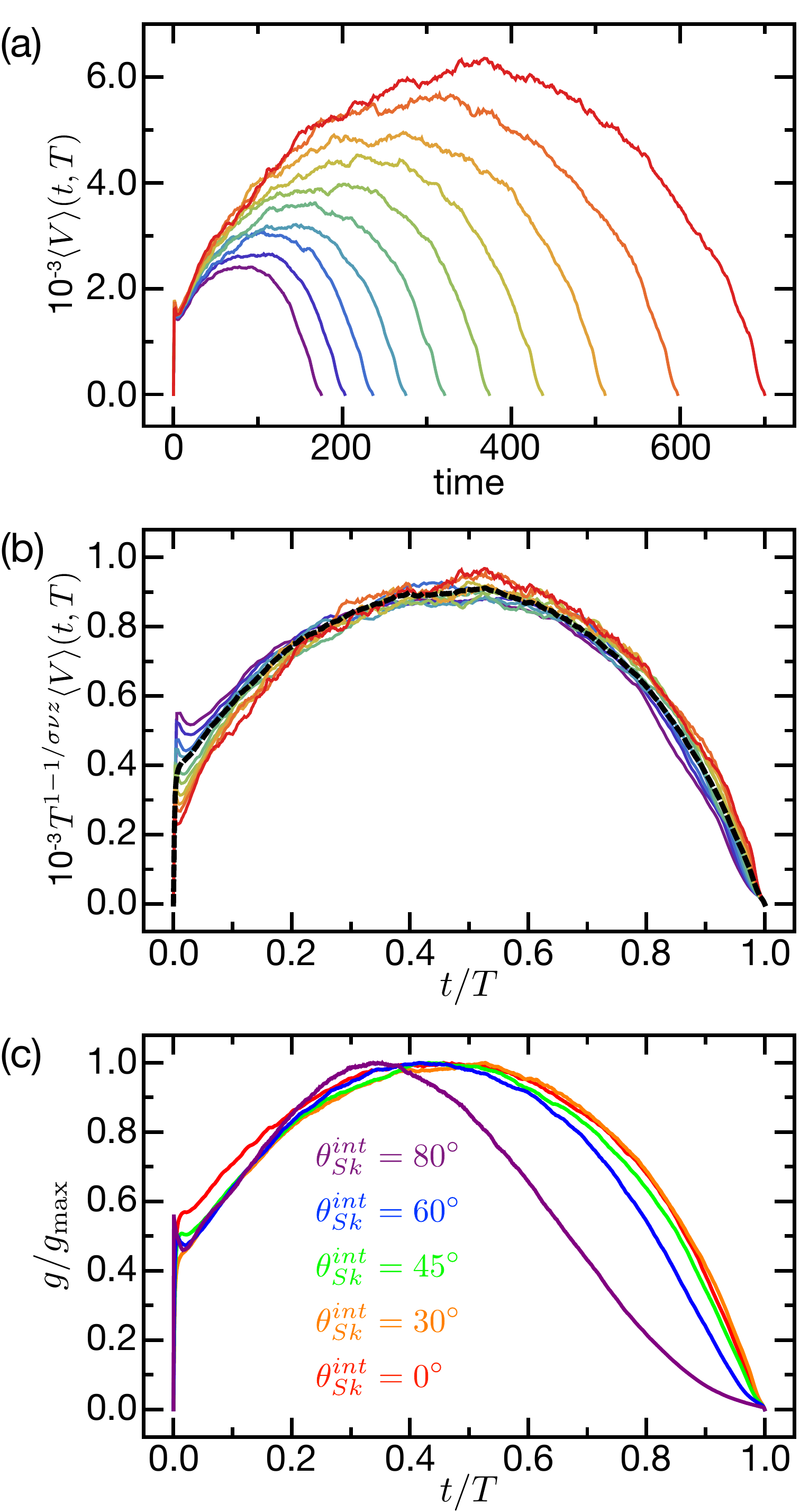}
\caption{(a)
  The time averaged avalanche  velocity $\langle V\rangle(t,T)$
  in a system with $\theta^{\rm int}_{Sk}=30^\circ$, for avalanches of duration $T$, versus time in simulation time steps. The curves represent the time average over ten logarithmically-spaced bins
  for $T = 150$, 175, 204, 238, 278, 324, 378, 441, 514, 600, and $700$
  simulation time steps, from left to right.
  (b) Scaling collapse of the data in panel (a) plotted as
  $T^{1 - 1/\sigma\nu z}\langle V\rangle(t,T)$ vs $t/T$,
  where  $1/\sigma\nu z = 1.63$.
  The dashed curve indicates the overall average avalanche shape.
  (c) The average avalanche shapes $g/g_{max}$ vs $t/T$ for
  $\theta^{\rm int}_{Sk} = 0^\circ$ (red), $30^{\circ}$ (orange), $45^{\circ}$ (green),
  $60^{\circ}$ (blue), and $80^{\circ}$ (purple).
}
\label{fig:3}
\end{figure}

A more stringent test of whether a system is at
a nonequilibrium critical point is
the prediction
that the average avalanche shape can be scaled to a universal curve \cite{31,32,33}. 
This
implies that the average skyrmion velocity for a given avalanche duration
should  scale as  $\langle V\rangle(t,T) \propto T^{1/\sigma\nu z - 1}g(t/T)$,
where $g(t/T)$ is a universal function of the avalanche shape
that can be extracted from the time
series by plotting $T^{1 - 1/\sigma\nu z}\langle V\rangle(t,T)$ versus $t/T$.    
In Fig.~\ref{fig:3}(a) we plot the average avalanche shape $\langle V\rangle$ for
different values of $T$ in the $\theta^{\rm int}_{Sk} = 30^\circ$ system, and in
Fig.~\ref{fig:3}(b) we show a
scaling collapse of the same data
versus $t/T$.
The dashed line is a fit to the overall average avalanche shape $g(t/T)$. 
We performed similar scaling collapses for other values
of $\theta^{\rm int}_{Sk}$ and find the same
universal function $g(t/T)$ for $\theta^{\rm int}_{Sk} \leq 60^{\circ}$, 
as shown in Fig.~\ref{fig:3}(c),
while for $\theta^{\rm int}_{Sk} = 80^{\circ}$,
the average avalanche shape is much more asymmetric.

\begin{figure}
\includegraphics[width=3.5in]{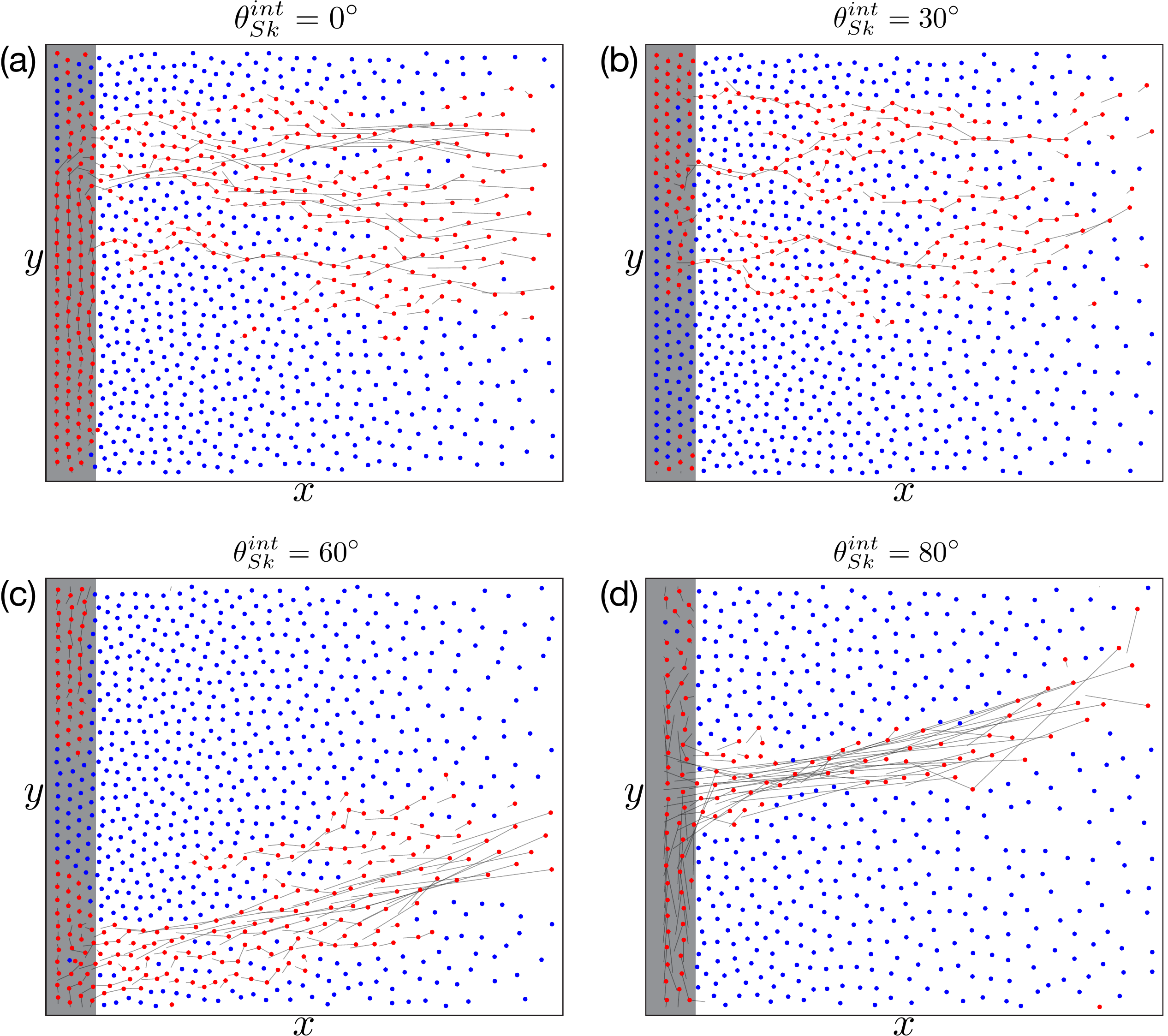}
\caption{Snapshots of the avalanche motion during a single large avalanche.  Blue dots
  indicate skyrmions that did not move during the avalanche event, red dots indicate
  skyrmions that moved a distance greater than $x$, and lines indicate the net
  displacement of individual skyrmions during the avalanche.
  (a) $\theta^{\rm int}_{Sk} = 0^\circ$. (b) $\theta^{\rm int}_{Sk} = 30^\circ$.
  (c) $\theta^{\rm int}_{Sk}=60^\circ$. (d) $\theta^{\rm int}_{Sk} = 80^\circ$.
  As the Magnus term increases,
  the avalanche motion starts to show curvature in the positive
$y$ direction.   
}
\label{fig:4}
\end{figure}

The change in the exponents and
the average avalanche shape
for large $\theta^{\rm int}_{Sk}$
indicates that when the non-dissipative Magus term
is strong, there is a change in the universality class.
Mean field 
predictions give $\tau = 1.5$, $\alpha = 2.0$, $1/\sigma\nu z = 2.0$,
and a parabolic universal function
for the average avalanche shape \cite{36,37}.
In our system, $\tau = 1.33$, $\alpha = 1.5$, $1/\sigma\nu z = 1.63$, 
and the universal function $g(t/T)$ has a parabolic shape
with some asymmetry at small $t/T$. 
Since we are working in two dimensions and the skyrmion interaction range
is finite,
it may be expected that our system would not match the mean field picture; however,
it is clear that when the Magnus term is large,
the avalanche dynamics show a pronounced change.
The asymmetry in the scaling collapse
of the avalanches is similar to that found in many systems
including magnetic domain avalanches,
where it was argued to
result from an
effective negative mass \cite{25}.
Inertial effects with positive mass tend to give a leftward asymmetry,
while an effective negative mass
damps the avalanches at later times and produces a rightward
asymmetry
\cite{25}. 
The Magnus term causes the skyrmions to move in the direction perpendicular to
the applied external force, and this could
reduce the overall avalanche motion in the forward direction at later times,
resulting in the skewed average avalanche shape.
In Fig.~\ref{fig:4}(a)  we plot the skyrmions and their
net displacements during a large avalanche in a sample with
$\theta^{\rm int}_{Sk} = 0^\circ$, where the motion strongly follows the density
gradient from left to right.
At $\theta^{\rm int}_{Sk}=30^\circ$ in Fig.~\ref{fig:4}(b),
near the right edge of the sample the avalanche motion shows
a tendency  to curve in the positive $y$ direction.
This tendency is enhanced for $\theta^{\rm int}_{Sk}=60^\circ$ in Fig.~\ref{fig:4}(c)
and for $\theta^{\rm int}_{Sk}=80^\circ$ in Fig.~\ref{fig:4}(d),
where the entire avalanche moves
at an angle with respect to the $x$ axis.   

We have examined several other pinning landscapes, including samples with
the same $N_p=3700$ but a lower $F_p=0.3$, where we find results similar to
those of the $F_p=1.0$ system.
In the limit of strong dilute pinning with $N_p=600$ and $F_p=3.0$, skyrmions that
become pinned generally never depin and we observe a strong channeling effect
where the avalanches occur through the motion of interstitial or unpinned skyrmions
moving along weak links between pinned skyrmions.  In this case, the distribution of
avalanche sizes is strongly peaked at the size corresponding to the weak link channel.

{\it Summary---}
We have shown that skyrmions driven by their own gradient
in the presence of quenched disorder exhibit avalanche dynamics and show
power law avalanche duration and size distributions.
The average avalanche 
shape for different avalanche durations
can be scaled by a universal function, in agreement with
predictions for systems near a nonequilibrium critical point.
Skyrmions are distinct from previously studied avalanche systems
due to the strong non-dissipative Magnus term in the skyrmion dynamics.
We find that as the Magnus term increases,
there is a change in the critical behavior of the avalanches as indicated by
the critical exponents, 
and the average avalanche shape develops a
strong asymmetry similar to that found for
a negative effective
mass in magnetic domain depinning avalanches.
This change in behavior
results when the
Magnus term causes the avalanche motion to shift partially into the direction
perpendicular to the skyrmion density gradient.

\acknowledgments
We thank Karin Dahmen for useful discussions. This work was carried out under the auspices of the 
NNSA of the U.S. DoE at LANL under Contract No.
DE-AC52-06NA25396.

\end{document}